\documentclass[DIV12]{scrartcl}

\usepackage{hyperref}
\usepackage[utf8]{inputenc}
\usepackage{tabularx}
\usepackage{setspace}
\usepackage{graphicx}
\usepackage{booktabs}
\usepackage{amsmath}
\usepackage[usenames,dvipsnames]{xcolor}

\usepackage{tikz}

\hypersetup{
	colorlinks=true,
	linkcolor=black,
	citecolor=black,
	urlcolor=black
}

\usepackage{subcaption}
\date{}
\usepackage{authblk}
\usepackage{blindtext}

\usepackage{natbib}
\usepackage{fancyhdr}
\title{Modeling Musical Context Using Word2vec}

\begin{document}

\author[1]{D. Herremans\thanks{d.herremans@qmul.ac.uk}}
\author[2]{C.-H. Chuan\thanks{c.chuan@unf.edu}}

\affil[1]{\small Queen Mary University of London, London, UK}
\affil[2]{\small University of North Florida, Jacksonville, USA}

\maketitle
\thispagestyle{fancy} 

\begin{abstract}

We present a semantic vector space model for capturing complex polyphonic musical context. A word2vec model based on a skip-gram representation with negative sampling was used to model slices of music from a dataset of Beethoven's piano sonatas. A visualization of the reduced vector space using t-distributed stochastic neighbor embedding shows that the resulting embedded vector space captures tonal relationships, even without any explicit information about the musical contents of the slices. Secondly, an excerpt of the Moonlight Sonata from Beethoven was altered by replacing slices based on context similarity. The resulting music shows that the selected slice based on similar word2vec context also has a relatively short tonal distance from the original slice.

\bigskip

\end{abstract}

\section{Introduction}
In this paper, we explore the semantic similarity that can be derived by looking solely at the context in which a musical slice appears. 
In past research, music has often been modeled through Recursive Neural Networks (RNNs) combined with Restricted Bolzmann Machines~\citep{boulanger2012modeling}, Long-Short Term RNN models~\citep{eck2002finding, sak2014long}, Markov models~\citep{conklin1995multiple} and other statistical models, using a representation that incorporates musical information (i.e., pitch, pitch class, duration, intervals, etc.). In this research, we focus on modeling the \emph{context}, over the content. 

Vector space models~\citep{rumelhart1988learning} are typically used in natural language processing (NLP) to represent (or embed) words in a continuous vector space~\citep{turney2010frequency, mcgregor2015distributional, agres2016modeling, liddy1999multilingual}. Within this space, semantically similar words are represented geographically close to each other~\citep{turney2010frequency}.  A recent very efficient approach to creating these vector spaces for natural language processing is word2vec~\citep{mikolov2013distributed}.

Although music is not the same as language, it possesses many of the same types of characteristics.  \citet{besson2001comparison} discuss the similarity of music and language in terms of, among others, structural aspects and the expectancy generated by both a word and a note. We can therefore use a model from NLP: word2vec. More specifically a skip-gram model with negative sampling is used to create and train a model that captures musical context.

There have been only few attempts at modeling \emph{musical} context with semantic vector space models. For example, \citet{huang2016chordripple} use word2vec to model chord sequences in order to recommend chords other than the `ordinary' to novice composers. In this paper, we aim to use word2vec for modeling musical context in a more generic way as opposed to a reduced representation as chord sequences. We represent complex polyphonic music as a sequence of equal-length slices without any additional processing for musical concepts such as beat, time signature, chord tones and etc. 

In the next sections we will first discuss the implemented word2vec model, followed by a discussion of how music was represented. Finally, the resulting model is evaluated. 







\section{Word2vec}

Word2vec refers to a group of models developed by \citet{mikolov2013distributed}. They are used to create and train semantic vector spaces, often consisting of several hundred dimensions, based on a corpus of text~\citep{mikolov2013efficient}. In this vector space, each word from the corpus is represented as a vector.  Words that share a context are geographically close to each other in this space. The word2vec architecture can be based on two approaches: a continuous bag-of-words, or a continuous skip-gram model (CBOW). The former uses the context to predict the current word, whereas the latter uses the current word to predict surrounding words~\citep{mikolov2013exploiting}. Both models have a low computational complexity, so they can easily handle a corpus with a size ranging in the billions of words in a matter of hours. While CBOW models are faster, it has been observed that skip-gram performs better on small datasets~\citep{mikolov2013efficient}. We therefore opted to work with the latter model.

\paragraph{Skip-gram with negative sampling} 

The architecture of a skip-gram model is represented in Figure~\ref{fig:skipgram}. For each word $w_t$ in a corpus of size $T$ at position $t$, the network tries to predict the surrounding words in a window $c$ ($c=2$ in the figure). The training objective is thus defined as: 

\begin{equation}
\frac{1}{T}\sum_{t=1}^T \sum_{-c \le i \le c, i \neq 0} \log p(w_{t+i} | w_t),
\end{equation}
whereby the term $p(w_{t+i} | w_t)$ is calculated by a softmax function. Calculating the gradient of this term is, however, computationally very expensive. Alternatives to circumvent this problem include hierarchical softmax~\citep{morin2005hierarchical} and noise contrastive estimation~\citep{gutmann2012noise}. The word2vec model used in this research implements a variant of the latter, namely negative sampling. 

\begin{figure}[h] \centering
\include{graph}
\caption{A skip-gram model with $n$-dimensions for word $w_t$ at position $t$. }
\label{fig:skipgram}
\end{figure}

The idea behind negative sampling is that a well trained model should be able to distinguish between data and noise~\citep{goldberg2014word2vec}. 
The original training objective is thus approximated by a new, more efficient, formulation that implements a binary logistic regression to classify between data and noise samples. When the model is able to assign high probabilities to real words and low probabilities to noise samples, the objective is optimized~\cite{mikolov2013distributed}.


\paragraph{Cosine similarity} was used as a similarity metric between two musical-slice vectors in our vector space. For two non-zero vectors A and B in $n$ dimensional space, with an angle $\theta$, it is defined as~\citep{Tan:2005}:

\begin{equation}
\text{Similarity(A, B)} = cos(\theta) = \frac{\sum_{i=1}^n{A_i \times B_i}}{\sum_{i=1}^n{A^2_i} \times \sum_{i=1}^n{B^2_i}}
\end{equation}

In this research, we port the above discussed model and techniques to the field of music. We do this by replacing `words' with `slices of polyphonic music'. The manner in which this is done is discussed in the next section.

\section{Musical slices as words}
\label{sec:slice}
In order to study the extend to which word2vec can model \emph{musical} context, polyphonic musical pieces are represented with as little injected musical knowledge as possible. Each piece is simply segmented into equal-length, non-overlapping slices. The duration of these slices is calculated for each piece based on the distribution of time between note onsets. The smallest amount of time between consecutive onsets that occurs in more than 5\% of all cases is selected as the slice-size. The slices capture all pitches that sound in a slice: those that have their onset in the slice, and those that are played and held over the slice. The slicing process does not depend on musical concepts such as beat or time signature; instead, it is completely data-driven. Our vocabulary of words, will thus consist of a collection of musical slices. 

In addition, we do not label pitches as chords. All sounding pitches, including chord tones, non-chord tones, and ornaments, are all recorded in the slice. We do not reduce pitches into pitch classes either, i.e., pitches C\textsubscript{4} and C\textsubscript{5} are considered different pitches. The only musical knowledge we use is the global key, as we transpose all pieces to either C major or A minor before segmentation. This enables the functional role of pitches in tonality to stay the same across compositions, which in turn causes there to be more repeated slices over the dataset and allows the model to be better trained on less data. 
In the next section, the performance of the resulting model is discussed.

\section{Results}


In order to evaluate how well the proposed model captures musical context, a few experiments were performed on a dataset consisting of Beethoven's piano sonatas. The resulting dataset consists of 70,305 words, with a total of 14,315 unique occurrences. As discussed above, word2vec models are very efficient to train. Within minutes, the model was trained on the CPU of a MacBook Pro. 

\begin{figure}[h!]
\centering
\begin{subfigure}[h]{.44\textwidth}
\includegraphics[width=\textwidth, height=6.4cm]{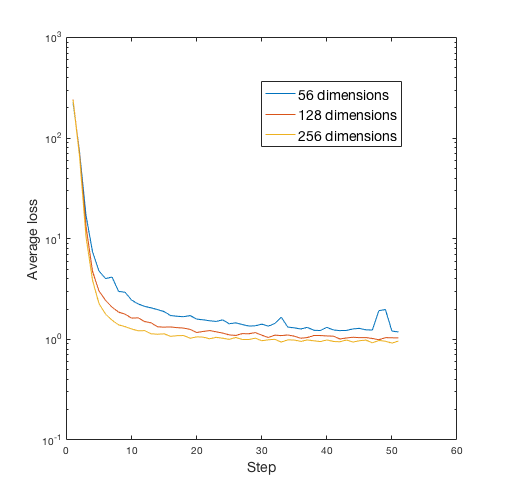}
\caption{Results for varying the number of dimensions of the vector space. }
\label{fig:dim}
\end{subfigure}
\begin{subfigure}[h]{.44\textwidth}
\includegraphics[width=\textwidth, height=6.4cm]{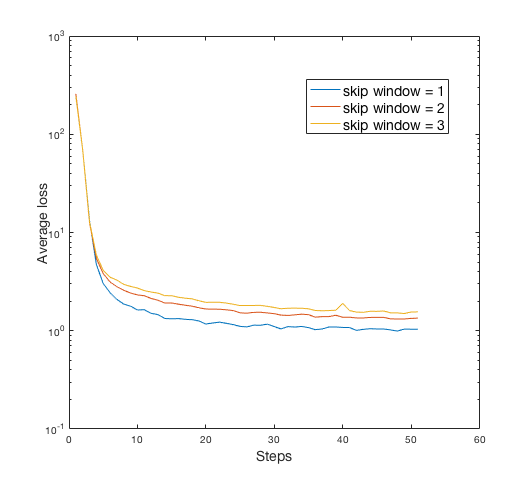}
\caption{Results for varying the size of the skip window. }
\label{fig:window}
\end{subfigure}
\caption{Evolution of the average loss during training. A step represents 2000 training windows. }
\end{figure}

We trained the model a number of times, with a different number of dimensions of the vector space (see Figure~\ref{fig:dim}). The more dimensions there are, the more accurate the model becomes, however, the time to train the model also becomes longer. In the rest of the experiments, we decided to use 128 dimensions.  In a second experiment, we varied the size of the skip window, i.e., how many words to consider to the left and right of the current word in the skip-gram. The results are displayed in Figure~\ref{fig:window}, and show that a skip window of 1 is most ideal for our dataset.

\subsection{Visualizing the semantic vector space}

In order to better understand and evaluate the proposed model, we created visualizations of selected musical slices in a dimensionally reduced space. We use t-Distributed Stochastic Neighbor Embedding (t-SNE), a technique developed by~\citet{maaten2008visualizing} for visualizing high-dimensional data. 
t-SNE has previously been used in a music analysis context for visualizing clusters of musical genres based on musical features~\citep{hamel2010learning}. 

\begin{figure}[h!] \centering
\begin{subfigure}[h]{.44\textwidth}
\includegraphics[width=\textwidth]{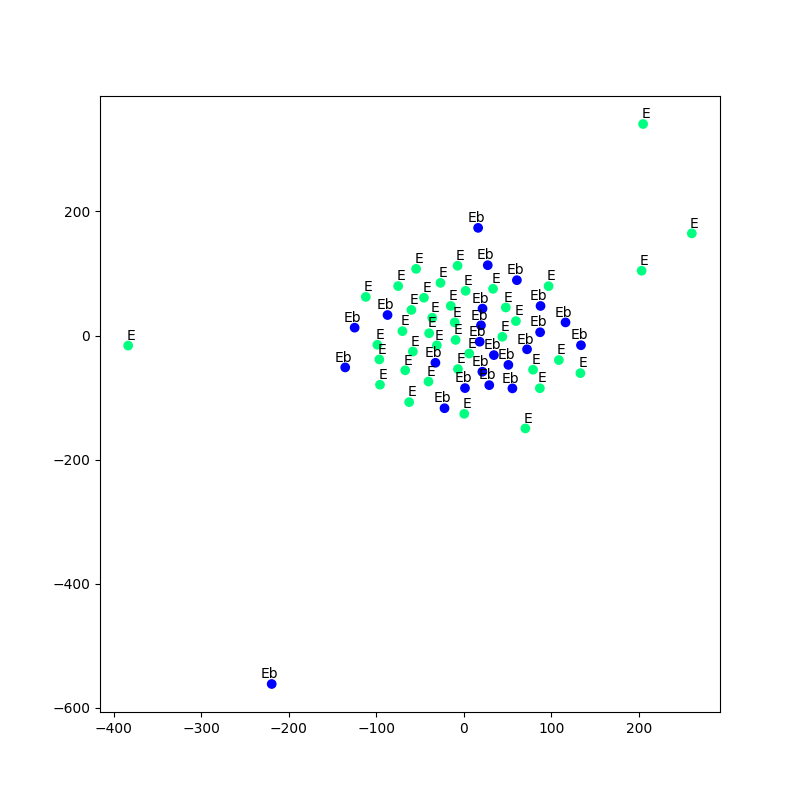}
\caption{E (green) and Eb (blue).}
\label{fig:vis1}
\end{subfigure}
\begin{subfigure}[h]{.44\textwidth}
\includegraphics[width=\textwidth]{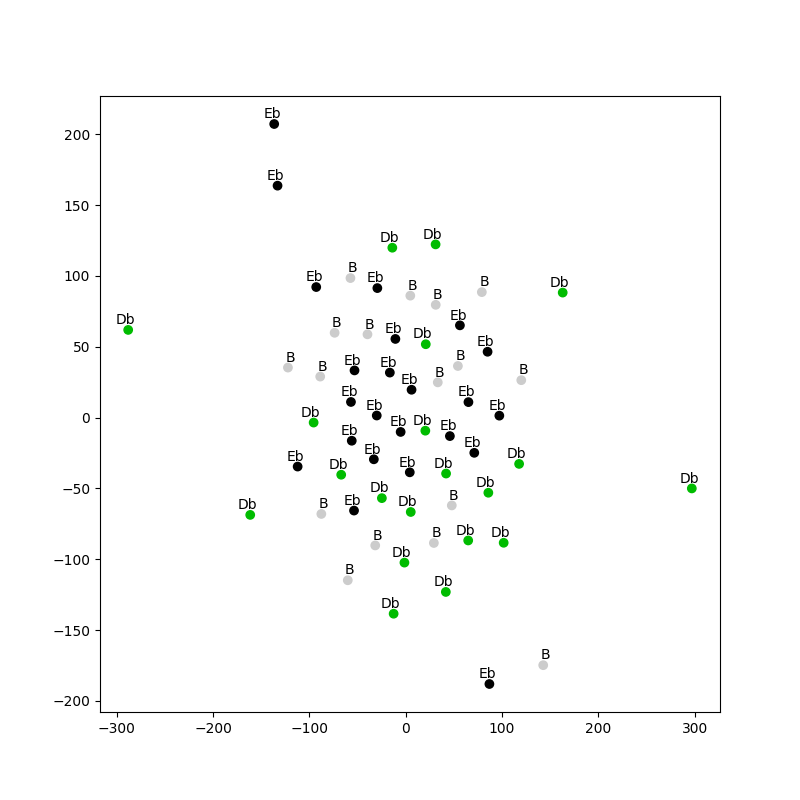}
\caption{Eb (black), Db (green) and B (gray).}
\label{fig:vis2}
\end{subfigure}
\begin{subfigure}[h]{.44\textwidth}
\includegraphics[width=\textwidth]{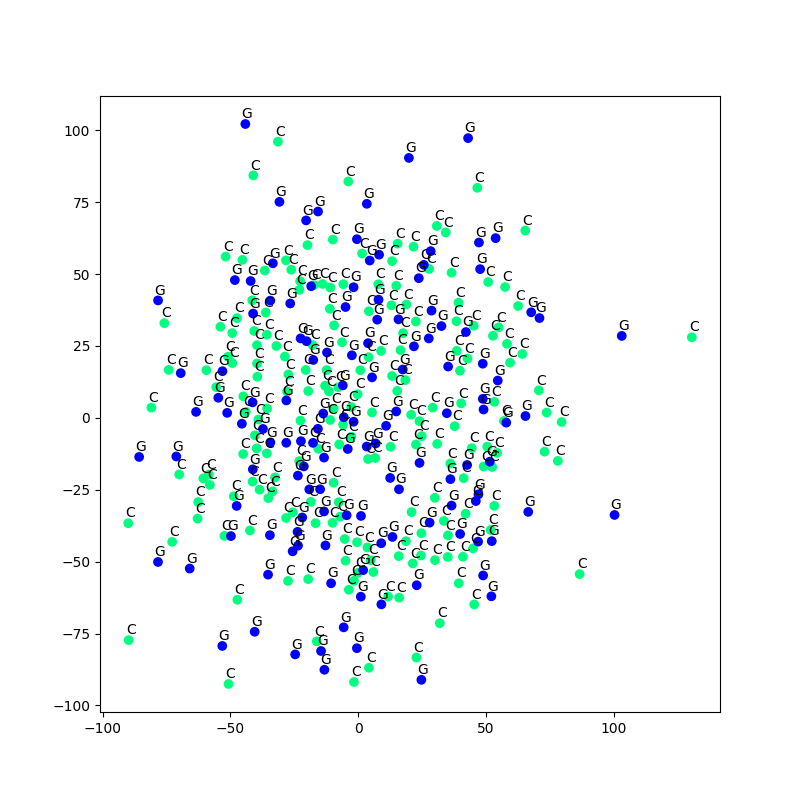}
\caption{C (green) and G (blue).}
\label{fig:vis3}
\end{subfigure}
\begin{subfigure}[h]{.44\textwidth}
\includegraphics[width=\textwidth]{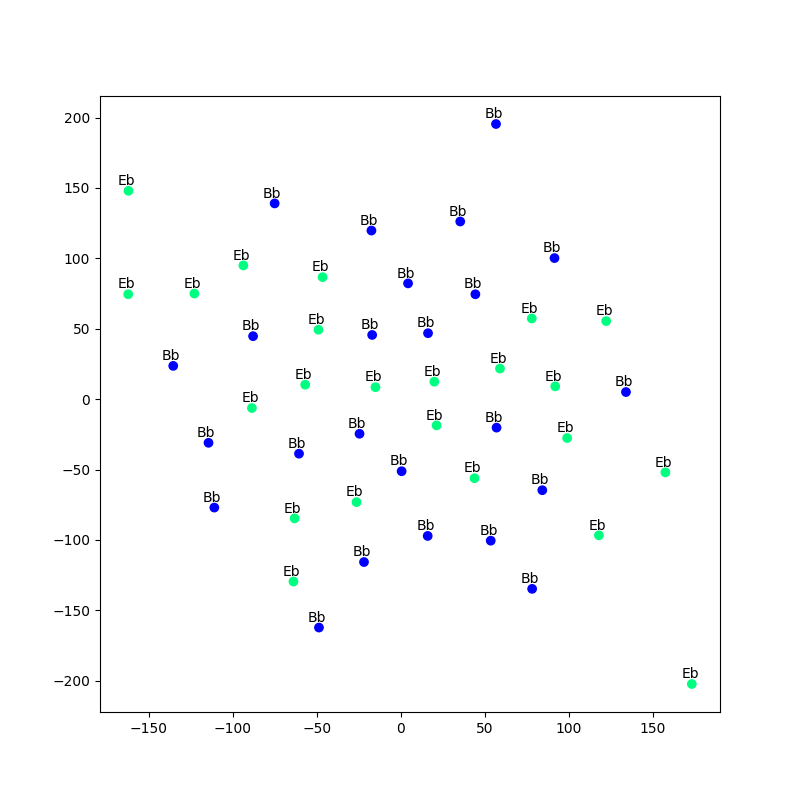}
\caption{Eb (green) and Bb (blue).}
\label{fig:vis4}
\end{subfigure}
\caption{Reduced vector space with t-SNE for different slices (labeled by the most close chord)}
\label{fig:vis}
\end{figure}

In this case, we identified the `chord' to which each slice of the dataset belongs based on a simple template-matching method. We expect that tonally close chords occur together in the semantic vector space. Figure~\ref{fig:vis} confirms this hypothesis. When examining slices that contain C and G chords (a perfect fifth apart), the space looks very dispersed, as they often co-occur (see Figure~\ref{fig:vis3}). The same occurs for the chord pair E\textsuperscript{b} and B\textsuperscript{b} in Figure~\ref{fig:vis4}. On the other hand, when looking at the tonally distant chord pair E and E\textsuperscript{b} (Figure~\ref{fig:vis1}), we see that clusters appear in the reduced vector space. The same happens for the tonally distant chords E\textsuperscript{b}, B\textsuperscript{b} and B in Figure~\ref{fig:vis2}.




\subsection{Content versus context}

In order to further examine if word2vec captures semantic meaning in music via the modeling of context, we modify a piece by replacing some of its original slices with the most similar one as captured by the cosine similarity in the vector space model. If word2vec is really able to capture this, the modified piece should sound similar to the original. This allows us to evaluate the effectiveness of using word2vec for modeling music. 

\begin{figure}[h]
\centering
\includegraphics[width=0.8\textwidth]{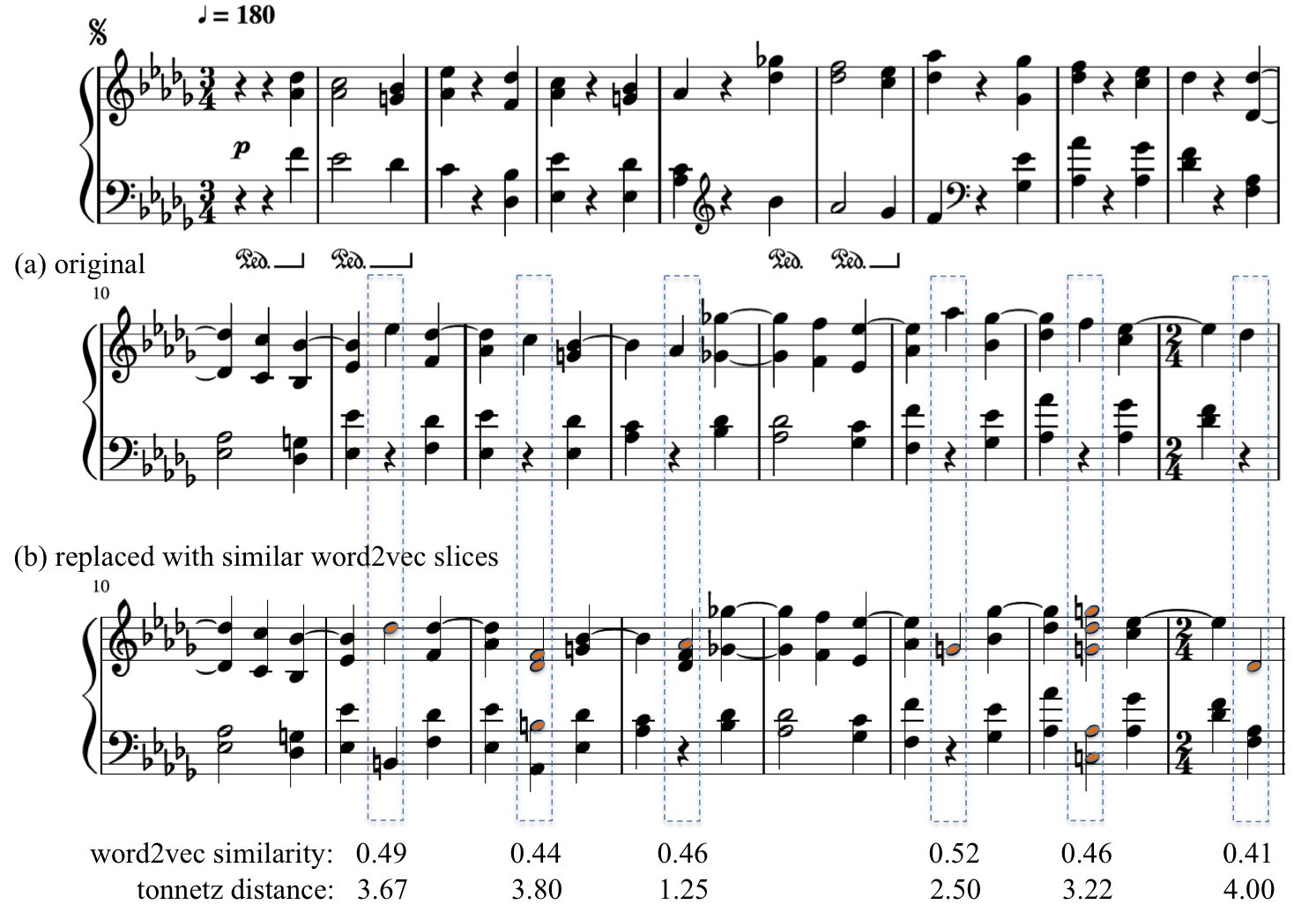}
\caption{(a) An excerpt of Beethoven's piano sonata Op. 27 No. 2, 2nd movement with (b) modified measures by replacing with slices that report the highest word2vec cosine similarity.}
\label{fig:generation}
\end{figure}

Figure~\ref{fig:generation} shows the first 17 measures of Beethoven's piano sonata Op. 27 No. 2 (Moonlight), 2nd movement in (a) and the measures with modified pitch slices in the dashed box in (b). An audio version of this score is available online\footnote{http://dorienherremans.com/word2vec}. The modified slices in (b) are produced by replacing the original with the slice that has the highest cosine similarity based on the word2vec embeddings. The tonal distance between the original and modified slices is presented below each slice pair.  This is calculated as the average of the number of steps between each pair of pitches in the two slices in a tonnetz representation~\citep{cohn1997neo}, extended with pitch register. It can be observed that even thought the cosine similarity is around 0.5, the tonal distance of the selected slice remains relatively low in most of the cases. For example, the tonal distance in the third dashed box between the modified slice (D\textsuperscript{b} major triad with pitches D\textsuperscript{b}\textsubscript{4}, F\textsubscript{4}, and A\textsuperscript{b}\textsubscript{4}) and the original slice of a single pitch A\textsuperscript{b}\textsubscript{4} is 1.25. However, we notice that word2vec does not necessarily model musical context for voice leading. For example, better voice leading can be achieved if the pitch D\textsubscript{4} in the last dashed box is replaced with pitch D\textsubscript{5}. 

In Figure~\ref{fig:generation}b, a number of notes are marked in a different color (orange). These are the held notes, i.e., their onsets are played previously and the notes remain being played over the current slice. These notes create a unique situation in music generation using word2vec. For example, the orange note (pitch D\textsuperscript{b}\textsubscript{5}) in the first dashed box is a held note, which indicates that the pitch should have been played in the previous slice. However, word2vec does not capture this relation; it only considers the similarity between the original and modified slices.




\section{Conclusions}

A skip-gram model with negative sampling was used to build a semantic vector space model for complex polyphonic music. By representing the resulting vector space in a reduced two-dimensional graph with t-SNE, we show that musical features such as a notion of tonal proximity are captured by the model. Music generated by replacing slices based on word2vec context similarity also presents close tonal distance compared to the original. 

In the future, an embedded model that combines both word2vec with, for instance, a long-short term memory recurrent neural network based on musical features, would offer a more complete way to more completely model music. The TensorFlow code used in this research is available online\footnote{http://dorienherremans.com/word2vec}.



\section*{Acknowledgements} 
This project has received funding from the European Union’s Horizon 2020 research and innovation programme under grant agreement No 658914.

\bibliography{paper}

\end{document}